\documentstyle[12pt]{article}
\begin{document}
\hoffset = -1truecm
\voffset = -2truecm
\title{\bf
Relation between space-time inversion and particle-antiparticle symmetry
and the microscopic essence of special relativity}

\author{
{\bf Guang-jiong Ni and Su-qing Chen}\thanks{gjni@fudan.ihep.ac.cn}\\ \\
{\it Physics Department, Fudan University, Shanghai, 200433, China\thanks{
Permanent address}}\\
 and \\
{\it
International Centre for Theoretical Physics, Trieste, Italy
 }}

\newpage

\maketitle

\begin{abstract}
  After analyzing the implication of investigations on the C, P and T
transformations since 1956, we propose that there is a basic symmetry in
particle physics. The combined space-time inversion is equivalent to
particle-antiparticle transformation, denoted by ${\cal PT=C}$. It is shown
that
the relativistic quantum  mechanics and quantum field theory do
contain this invariance explicitly or implicitly. In
particular, (a) the appearance of negative energy or negative
probability density in single particle theory -- corresponding
to the fact of existence of antiparticle, (b) spin- statistics
connection, (c) CPT theorem, (d) the Feynman propagator are
linked together via this symmetry. Furthermore, we try to
derive the main results of special relativity, especially, (e) the
mass-energy relation, (f) the Lorentz transformation by this one
``relativistic" postulate and some ``nonrelativistic" knowledge.
 \end{abstract}

\newpage
\leftline{{\Large \bf I. INTRODUCTION}}
\vskip.2in
  Since the historic discovery of parity violation in week interactions by
Lee and Yang [1] and the experimental verification by Wu et al. [2], the
investigation on C,P and T problems has been attracting serious attention
in physics community. In 1964, Christensen et al. discovered the CP
violation in the decay process of neutral K meson [3]. The further
analysis showed that the T inversion is certainly violated whereas the
CPT theorem still remains valid [4].

  The purpose of this paper is trying to examine this problem from an
alternative point of view [5]. In section II  we suggest that it is the
time to propose a new principle (postulate) as the replacement of CPT
theorem. Then in sec.III the Dirac equation is analyzed in detail from the
new point of view, the Klein-Gordon equation and Maxwell equation are
also discussed. The sec. IV is devoting to discussing the connection
between spin and statistics. In sec. V we make an observation on the
Feynman propagator and the arrow of time in physics which are intimately
related to this basic symmetry under consideration. Then in sec. VI we will
be able to derive the main results of special relativity by means of this
symmetry with some other "nonrelativistic" knowledge. The final section VII
contains brief summary and discussions. Some other details are given in
three appendices.
          \vskip.2in
\leftline{{\Large \bf II. WHERE IS THE PROBLEM?}  }
          \vskip.2in
  The discovery of parity violation led directly to the establishment of two
component neutrino theory [6]
$$
|\bar{\nu}> = \mbox{\tt CP} |\nu>   \eqno(2.1)
$$

    The success of this theory implies the ineffectiveness of the original
definition of space reflection P and C transformation respectively. After
the discovery of CP violation, Lee and Wu proposed a unified definition
for particle-antiparticle transformation [7]:
$$
|\bar{a}>=\mbox{\tt CPT}|a>  \eqno(2.2)
$$

  This is really a very important process of evolution in concept. The
physicists have been correcting a long existing mistake in physics, the
latter conceives that the matter is only located in the space-time and
can be detached from it.

Here two remarks are important to us. (a). The difference between a
theorem and a law should be treated carefully. Every quantity in a
theorem must be defined unambiguously and separately before the theorem
can be proved. Actually, the conclusion of a deduction method is already
contained implicitly in the premise. On the other hand, since a law is
verified by experiments, it is often (not always) capable of accommodating
a definition of physical quantity, which is not defined independently
before the law is established. (b).
As we learn from the development of physics in this century,
the definition of any observable in physics must depends on some
invariance or dynamical law. Once it fails to do so, it will cease to be
an observable. For example, the energy $E$
 (momentum $\stackrel{\rightarrow}{p}$) can be defined
because of the existence of law of energy (momentum) conservation. The
definition of inertial mass $m$ in Newtonian mechanics is contained in the
law $\stackrel{\rightarrow}{F}=m\stackrel{\rightarrow}{a}$.
 However, in the theory of special relativity, $m$ should be
defined as $m=d\stackrel{\rightarrow}{p}^2/2dE$ for taking the changeableness
of mass into
account (see Eq. (6.14) below). Therefore, some time the change of
definition is necessary and important. To some extent, this is also true
for the definition of a transformation, e.g. ,the particle- antiparticle
transformation.

  The ineffectiveness of individual definitions of P, T and C together
with the validity of CPT theorem [8, 9] enlightened us that one should
introduce new definition of space-time inversion and replace the CPT
theorem by a fundamental principle (postulate) which can be stated as
follows:

  Under the combined space-time inversion, all particles with mutual
interactions turn to their antiparticles respectively.

  Here the meaning of inversion needs to be clarified. First consider the
single particle quantum mechanics. The space reflection operator is
denoted by ${\cal P}$. There are two equivalent statements:

(A) If there is a physical state described by wave function
$\psi(\stackrel{\rightarrow}{x},t)$ in the
coordinate system $\{x\}$, then after substitution
$\stackrel{\rightarrow}{x}=-\stackrel{\rightarrow}{x'}$, the wave function
changes to that in reversed system $\{x'\}$ as follows:
$$
\psi(\stackrel{\rightarrow}{x},t)\longrightarrow\psi(-\stackrel{\rightarrow}{x'},t)=\psi'(\stackrel{\rightarrow}{x'},t'),\ (t=t')  \eqno(2.3)
$$
The substitution $\stackrel{\rightarrow}{x}=-\stackrel{\rightarrow}{x'}$ also
has to be made in the equations.

(B) Instead of introducing reversed system $\{x'\}$, one may introduce the
space reflected state in the same $\{x\}$
system according to the following rule:
$$
\psi(\stackrel{\rightarrow}{x},t)\longrightarrow\psi(-\stackrel{\rightarrow}{x},t)=\psi'(\stackrel{\rightarrow}{x},t),  \eqno(2.4)
$$
The corresponding change $\stackrel{\rightarrow}{x}\rightarrow
-\stackrel{\rightarrow}{x}$ also has to be made in the
equations.

  We shall adopt (A) or (B) statement freely in the later discussion.

  Similarly, if adopting (B) statement, the time reversal operator ${\cal T}$
means
that
$$
t \longrightarrow -t,\
\psi(\stackrel{\rightarrow}{x},t)\longrightarrow
\psi(\stackrel{\rightarrow}{x},-t)\ .
\eqno(2.5) $$

  Notice that, however, we do not demand that the physical law is invariant
under the ${\cal P}$ or ${\cal T}$ inversion individually. In other words,
whether the
space or time reflected state in the right hand side of (2.4) or (2.5)
exists or not, a concrete analysis for different situation is needed.

But the fundamental postulate just mentioned before claims that:
During
$$ \stackrel{\rightarrow}{x}\longrightarrow -\stackrel{\rightarrow}{x},\
t\longrightarrow -t\ ,$$
$$
\psi(\stackrel{\rightarrow}{x},t)\longrightarrow\psi(-\stackrel{\rightarrow}{x},-t)=\psi_c(\stackrel{\rightarrow}{x},t),  \eqno(2.6)
$$
where the space-time reflected state in the right hand side,
$\psi_c(\stackrel{\rightarrow}{x},t)$, is
just the antiparticle state corresponding to
 $\psi(\stackrel{\rightarrow}{x},\ t)$, i. e. ,
$$
  \psi_c(\stackrel{\rightarrow}{x},t)={\cal
C}\psi(\stackrel{\rightarrow}{x},t)\ . \eqno(2.7)
$$
Here we introduce a new particle- antiparticle conjugate operator ${\cal C}$,
which is not defined independently because the postulate (2.6) implies that
$$ {\cal PT}={\cal C} \ . \eqno(2.8) $$

  In quantum mechanics, the momentum and energy operators for a particle
read:
$$
\hat{\stackrel{\rightarrow}{p}}=-i \hbar \nabla\ ,\
\hat{E}=i\hbar\frac{\partial}{\partial t} \ .  \eqno(2.9)
$$
Note that after the space-time inversion, the components of four
momenta of the antiparticle remain unaltered, i.e.,
(We adopt the Pauli metric. For notation see Ref. [10])

  $$ p^c_{\mu}=p_{\mu}\ \ (\mu=1,\ 2,\ 3,\ 4)\ ,$$
while
 $$
 p_{\mu}^c=<\psi_c|\hat p_{c\mu}|\psi_c>,
\ p_{\mu}=<\psi|\hat p_{\mu}|\psi>\ .   \eqno(2.10)
$$

  Then from (2.6), we find that the momentum and energy operators for an
antiparticle should be:
$$
\hat{\stackrel{\rightarrow}{p}_c}=i \hbar \nabla\ ,\
\hat{E_c}=-i\hbar\frac{\partial}{\partial t} \ .
\eqno(2.11) $$
For example, the plane waves
$$
\psi(\stackrel{\rightarrow}{x},t)=\exp\{i(\stackrel{\rightarrow}{p}\cdot
\stackrel{\rightarrow}{x}-Et)/\hbar\}   \eqno(2.12)
$$
and
$$
\psi_c(\stackrel{\rightarrow}{x},t)=\exp\{-i(\stackrel{\rightarrow}{p}\cdot
\stackrel{\rightarrow}{x}-Et)/\hbar\}   \eqno(2.13)
$$
describe a particle and an antiparticle with the same momentum
$\stackrel{\rightarrow}{p}$ and
positive energy $E$ respectively. thus the phase of wave function is
important. From now on, if we meet a "negative energy" ($
\psi\sim \exp\{(iEt/\hbar)\},\ E>0$) in
the exponential, we should recognize it describing an antiparticle. This
point had been emphasized by Schwinger[11] and the expression (2.11) had
also been written down by Konopinski and Mahmaud[12]. But here (2.11) and
(2.13) are all the direct corollaries of our fundamental postulate. (From
now on the superscript of an operator, $\hat { }$, will be omitted).

  Comparing (2.2) and (2.7), we find that
$$
{\cal C}={\cal PT}=\mbox{\tt CPT} .\ \eqno(2.14)
$$

  In the following sections we will see that the present particle theory
does have such a corresponding relation. This is not accidental and is
not merely a change in definition. (See the discussion after Eq.(3.4) ).

  When considering the many-body problem, according to the quantum field
theory, the wave function of different kind of field is promoted into the
field operator in Hilbert space. Because the time inversion not only
renders the individual field operator to change its argument from
$t\rightarrow
-t$, but also reverses the order of field operators in a product. Hence
under the combined space-time inversion (denoted by $x\rightarrow  -x$),we
have
$$
\psi_1(x_1)
\psi_2(x_2)
\cdots
\psi_n(x_n)
\longrightarrow\psi_n'(-x_n)
\cdots
\psi_2'(-x_2)
\psi_1'(-x_1)\ ,
\eqno(2.15)
$$
where the superscript prime ``," is added for taking care the change of
annihilation and creation operators, (see sec. IV).

  According to this new definition, the configuration space and Hilbert
space are mixed together, so we do not use two kinds of notation of
space-time inversion for single particle and many particle theory
separately, but simply denote them by ${\cal P,\ T}$,  and ${\cal
C}={\cal PT}$.

  The CPT theorem was proved by Pauli and L\"uders [8] according to
the original definitions and making use of so called strong reflection
invariance. Thanks to CPT theorem and the researches over ten years, the
physicists arrive at the correct cognition (2.2). However, in our point
of view, once the relation (2.2)is reached, the historic mission of CPT
theorem is coming to an end. It should be replaced by a fundamental
postulate as shown in (2.8) [i.e. , (2.14)]. In other words, the relation
between space-time inversion and particle- antiparticle conjugation is
not a problem of definition and a proof of theorem, but a natural law
which anyhow must be tested by experiments. In this sense, the present
particle theory which is verified by numerous experimental facts will be
the basis of this postulate. On the other hand, one may get some new
insight from it.
\vskip.2in
\leftline{{\Large \bf III. THE RELATIVISTIC PARTICLE EQUATIONS}}
\vskip.2in
\leftline{{\large \bf A. The particle with spin 1/2}}
\vskip.2in
   The single particle theory basing on the Dirac equation is asymmetric
with respect to electron and positron. One has to overcome the negative
energy difficulty by means of the hole concept, then throw away the hole
by the method of redefinition when performing the second quantization in
order to obtain the formal symmetry between electron and positron
[10,13]. In our point of view, as the equal existence of electron and
positron is a fact in nature beyond any doubt, our theory should reflect
this symmetry at every step. Let us start from the new postulate and view
the negative energy solution directly as the wave function of positron.
Hence the Dirac equation [10]
$$
\left(\frac{\partial}{\partial x_{\mu}}-\frac{ie}{\hbar
c}A_{\mu}(\stackrel{\rightarrow}{x},\ t)\right)
\gamma_{\mu}\psi(\stackrel{\rightarrow}{x},\
t)+\frac{mc}{\hbar}\psi(\stackrel{\rightarrow}{x},\ t)=0
\eqno(3.1)
$$
($e<0,\ \gamma_k=-i\beta\alpha_k,\ \gamma_4=\beta$) not only
describes the electron, but also describes
the positron. In the "positive energy solution" which describes the
electron, the first and second components of spinor are large components
whereas the third and fourth components are small ones. On the other
hand, in the "negative energy solution", i.e., the positron wave function,
the large components of spinor go down to the third and fourth
position (As to the problem of spin orientation, see Appendix A). Hence
we may say that Dirac equation is mainly for describing the electron. Is
there an equivalent equation mainly for describing the position?

  Let us perform a ${\cal PT}$ inversion for Eq. (3.1):
$$
x_{\mu}\longrightarrow
-x_{\mu}\ ,\
\psi(x_{\mu})
\longrightarrow\psi(-x_{\mu})
=\psi_c(x_{\mu})
\eqno(3.2)$$
and note that at the same time electromagnetic potential will transform
as follows:
$$
A_{\mu}(\stackrel{\rightarrow}{x},\ t)
\longrightarrow A_{\mu}(-\stackrel{\rightarrow}{x},\ -t)
=-A_{\mu}(\stackrel{\rightarrow}{x},\ t)
\eqno(3.3)$$
This is because when we adhere to the fundamental postulate, not only the
electron under observation, but all the charges which create the
electromagnetic fields change their sign of charge as well. Thus we have
$$
\left(\frac{\partial}{\partial x_{\mu}}-\frac{ie}{\hbar
c}A_{\mu}(\stackrel{\rightarrow}{x},\ t)\right)
\gamma_{\mu}\psi_c(\stackrel{\rightarrow}{x},\
t)-\frac{mc}{\hbar}\psi_c(\stackrel{\rightarrow}{x},\ t)=0
\eqno(3.4)
$$
which corresponds to the transformation $m\rightarrow -m$ in (3.1) without
the change
of $e$, also corresponds to the representation transform
$\psi=\gamma_5\psi_c$, (chirality
transformation). In the past, a combined CPT transformation will lead to
CPT$=-\gamma_5$   if suitable phase is chosen [13].  Therefore, for a particle
with spin 1/2 we have illustrated Eq. (2.14).

  To our knowledge, Tiomno first noticed the equivalence of the Dirac
equation under $\gamma_5$   transformation, Sakurai had written down the Eq.
(3.4)
in 1958, Nambu and Jona-Lasinio in 1961 had derived a more general form:
[14]
$$
\left(\gamma_{\mu}\frac{\partial}{\partial x_{\mu}}+\frac{mc}{\hbar }
(\cos 2\alpha +i\gamma_5 \sin 2\alpha)\right)
\psi(x_{\mu})=0
\eqno(3.5)
$$

  However, the meaning of either (3.4) or (3.5) had not been clarified
before, so they did not get enough emphasis. In our point of view, being
the space-time inversed equation of Dirac equation, Eq. (3.4) describes
mainly the positron, but the electron as well. There is a simple relation
under ${\cal PT}$ inversion between its solution and the solution of
Dirac equation. This relation has nothing to do with the $\gamma$ matrices (see
Appendix B). For this reason we suggested to name Eq. (3.4) as the Carid
equation.

 Comparing the
$\psi_{III}(x)$ and $\psi_{cIII}(x)$
in Appendix B, we see that the relation
under space-time inversion is precisely: $\psi_{III}(x)$=$\psi_{cIII}(x')$
$ (x'=-x)$. In other words, a positron in $\{x\}$ system is just equivalent to
an electron (not a
negative-energy electron) in inversed $\{x'\}$ system. This implies a
modification to the St\"ueckelberg-Feynman rule [15] and is also a
special statement of fundamental postulate in this paper.

  Therefore, it seems to us that the original positron equation obtained
via $e\rightarrow -e$ transformation is incorrect. Positron and electron
obey the same equation$-$Dirac equation or Carid equation, the latter is
obtained
from the former via $m\rightarrow -m$ transformation. It is interesting to
compare with the classical theory, where the motion of equation for electron
$$
e\left(\stackrel{\rightarrow}{E}+\frac1c \stackrel{\rightarrow}{v}\times
\stackrel{\rightarrow}{B}\right)=m\frac{d \stackrel{\rightarrow}{v}}{d t}
\eqno(3.6)
$$
will lead to the motion equation for positron via either
$e\rightarrow -e$ or $m\rightarrow -m$
transformation. However, we stress here that the difference between
particle and antiparticle is not due to the difference in some "charge",
but due to their opposite space-time phases in their wave functions. See
also the Foldy-Wouthuysen transformation[13].

  As in Ref. [13], let us write
$$
\psi(x)=\left(\begin{array}{c} \theta(x) \\ \chi(x) \end{array}\right)
,\
\psi_c(x)=\left(\begin{array}{c} \chi_c(x) \\ \theta_c(x) \end{array}\right)
\eqno(3.7)
$$
such that the Dirac equation is resolved into two simultaneous equations
of two component spinors $\theta (x)$ and $\chi (x)$:

$$
\left\{ \begin{array}{l}
i\hbar\displaystyle{\frac{\partial \theta}{\partial t} =c
\stackrel{\rightarrow}{\sigma}\cdot
\left(\frac{\hbar}{i}\nabla-\frac{e}{c}\stackrel{\rightarrow}{A}\right)\chi(x)}+(eV+mc^2)\theta(x)
\cr
i\hbar\displaystyle{\frac{\partial \chi}{\partial t}= c
\stackrel{\rightarrow}{\sigma}\cdot
\left(\frac{\hbar}{i}\nabla-
\frac{e}{c}\stackrel{\rightarrow}{A}\right)\theta(x)}+(eV-mc^2)\chi(x)
\end{array}\right. \ .     \eqno(3.8)
$$

  Things become more symmetric
in new point of view. For an electron, $|\theta|>|\chi|,$ the $\theta (x)$
  characterizing electron
determines the phase of
space-time, i.e.,
$\chi\sim\theta \sim\exp (-iEt/\hbar),$ $  (E>  0)$.
For a
positron, the situation is
just in the opposite, $|\chi|>|\theta|      ,\
\theta\sim\chi\sim\exp (iEt/\hbar),$ $  (E>  0)$.
When
performing a space-time inversion to a electron wave function,
$\theta(\stackrel{\rightarrow}{x},\ t)
\rightarrow\theta(-\stackrel{\rightarrow}{x},\ -t)
=\chi_c(\stackrel{\rightarrow}{x},\ t)$
remains as large component, whereas
$\chi(\stackrel{\rightarrow}{x},\ t)\rightarrow
\chi(-\stackrel{\rightarrow}{x},\ -t)=
\theta_c(\stackrel{\rightarrow}{x},\ t)$
       remains small.
Hence an electron changes into a positron. Therefore, in some sense we
can say that an electron contains some ingredient of positron implicitly
and coherently. On the contrary, a positron state contains some
ingredient of electron implicitly and coherently too. Any discrimination
between particle from antiparticle is relative. There is no existence of
either pure particle or pure antiparticle even at the single particle
level. For further discussion see Appendix C.
\vskip.2in
\leftline{\large \bf B. The charged particle without spin}
\vskip.2in
   The Klein-Gordon equation describes a particle with spin zero:
$$
\left(\frac{1}{c^2}\frac{\partial^2}{\partial t^2}-\nabla^2
+\frac{m^2c^2}{\hbar^2}\right)
\phi(\stackrel{\rightarrow}{x},\ t)=0\ .    \eqno(3.9)
$$
Obviously, K-G equation remains unaltered under ${\cal PT}$ inversion,
whereas its complex wave solution does undergo the following change,

$$
\phi(\stackrel{\rightarrow}{x},\
t)\longrightarrow\phi(-\stackrel{\rightarrow}{x},\ -t)=
\phi_c(\stackrel{\rightarrow}{x},\ t)   \eqno(3.10) $$
which denotes a meson (say $\pi^-$ ) changes into antimeson (say
$\pi^+$    ).

Furthermore, the K-G equation with external electromagnetic potential can be
recast into two simultaneous equations of one order [16]. Denote

$$
D_{\mu}=\frac{\partial}{\partial x_{\mu}}-\frac{ie}{\hbar c}A_{\mu}
\eqno(3.11)
 $$
and

$$
\left\{\begin{array}{l}
-\frac{1}{\kappa}D_4\phi=\theta-\chi\ ,\ (\kappa=\frac{mc}{\hbar})
\cr
\phi=\theta+\chi
\end{array}\right.
\eqno(3.12)
$$
Instead of $\phi$, now one has two parts, $\theta$ and $\chi$, in coupling.
Then K-G equation reads

$$
\left\{ \begin{array}{l}
i\hbar\displaystyle{\frac{\partial \theta}{\partial t}=\frac{1}{2m}
\left(\frac{\hbar}{i}\nabla-\frac{e}{c}\stackrel{\rightarrow}{A}
\right)^2(\theta+\chi)}+
(eV+mc^2)\theta
\cr
i\hbar\displaystyle{\frac{\partial \chi}{\partial t}=-\frac{1}{2m}
\left(\frac{\hbar}{i}\nabla-\frac{e}{c}\stackrel{\rightarrow}{A}
\right)^2(\theta+\chi)}
+(eV-mc^2)\chi \end{array}\right. \ .     \eqno(3.13)
$$
Similar to Eq. (3.8), this time we also see that under the ${\cal PT}$
inversion:
$$
\theta(\stackrel{\rightarrow}{x},\ t)\longrightarrow
\theta(-\stackrel{\rightarrow}{x},\ -t)
=\chi_c(\stackrel{\rightarrow}{x},\ t),\
\chi(\stackrel{\rightarrow}{x},\ t)\longrightarrow
\chi(-\stackrel{\rightarrow}{x},\ -t)=
\theta_c(\stackrel{\rightarrow}{x},\ t)\ .
$$
If the former, $\theta$, is the larger one and so dominates the latter,
$\chi$,
then a particle in turn changes into antiparticle. Moreover, because the
probability density [3, 17]

$$
\rho=\frac{i\hbar}{2mc^2}\left(
\phi^*\frac{\partial\phi}{\partial t}
-\phi\frac{\partial\phi^*}{\partial t}
\right)  \eqno(3.14)
$$
changes its sign under ${\cal PT}$   inversion $(\phi\rightarrow \phi_c  )$, as
long
as we interpret   $\rho$
as a "charge density" [18], the so called "negative probability
difficulty" does not exist even at the level of single particle theory.

\vskip.2in
{\large \bf C. The photon}
\vskip.2in

   The Maxwell equation in vacuum can be recast in the following form
[19]:
$$
i\hbar\frac{\partial}{\partial t}\Phi=-i\hbar
c\stackrel{\rightarrow}{S}\cdot\nabla\Phi\ ,    \eqno(3.15) $$
where the vector operator $\stackrel{\rightarrow}{S}$ is $3\times3$
hermitian matrices with three components:
$$
S_1=i\left(\begin{array}{ccc}
0  &   0   &   0  \cr
0  &   0   &   -1  \cr
0  &   1   &   0
\end{array}\right)\ ,\
S_2=i\left(\begin{array}{ccc}
0  &   0   &   1  \cr
0  &   0   &   0  \cr
-1  &   0   &   0
\end{array}\right)\ ,\
S_3=i\left(\begin{array}{ccc}
0  &   -1   &   0  \cr
1  &   0   &   0  \cr
0  &   0  &   0
\end{array}\right)\ ,\
\eqno(3.16)
$$
while
$$\Phi(\stackrel{\rightarrow}{x},\ t)=\left(\begin{array}{c}
E_1+iB_1 \cr
E_2+iB_2 \cr
E_3+iB_3
\end{array}\right)\ ,\
\eqno(3.17)
$$
being the electromagnetic field "wave function". As before, it is
composed of two parts, $ \stackrel{\rightarrow}{E}$ and
$\stackrel{\rightarrow}{B}$. If introducing the "orbital angular
momentum" $\stackrel{\rightarrow}{L}$, and using the similar method in App. A,
one can easily
prove that:
$$
\frac{d}{d t}(\stackrel{\rightarrow}{L}+\hbar \stackrel{\rightarrow}{S})=0
\eqno(3.18)
$$
This implies that a "photon" has a spin angular momentum  $\hbar
\stackrel{\rightarrow}{S}$ with spin
quantum number $S=1$. Eq. (3.15) can also be written as
$$
H\Phi=c\stackrel{\rightarrow}{S}\cdot\stackrel{\rightarrow}{p}\Phi
\eqno(3.19)
$$
which bears a close resemblance to the Weyl equation (see Eqs. (6.18),
(6.19)). By means of the unified method in this paper, we see that there
are only two basic states of electromagnetic wave, i.e., the right and
left circular polarization states, (similar conclusion was arrived at by
Hestenes via another method [20]). The corresponding photons are
denoted as $\gamma_R$  and $\gamma_L$:
$$
|\gamma_L>={\cal PT}|\gamma_R>={\cal C}|\gamma_R>\ .
\eqno(3.20)
$$
Note that, however, we have

$$
\left\{\begin{array}{l}
{\cal P}|\gamma_R>=|\gamma_L>\ ,
\cr
{\cal T}|\gamma_R>=|\gamma_R>\ ,
{\cal T}|\gamma_L>=|\gamma_L>\ .\end{array}\right.
\eqno(3.21)
$$

\vskip.2in
\leftline{{\Large \bf IV. CONNECTION BETWEEN SPIN}}
\vskip.2in
\leftline{{\Large \bf AND STATISTICS}}
\vskip.2in
  In quantum field theory, the wave function becomes a field operator. In
this case we further assume that under ${\cal PT}$ inversion an
annihilation
operator $a(\stackrel{\rightarrow}{k})$ of particle with momentum
$\stackrel{\rightarrow}{k}$ and energy $E>0$ will change
to a
creation operator $b^{\dag}(\stackrel{\rightarrow}{k})$ of antiparticle with
momentum $\stackrel{\rightarrow}{k}$ and
energy $E> 0$. Thus we see that the complex scalar field operator
$$
\phi(x)=
\frac1{\sqrt{V}}\sum_{\stackrel{\rightarrow}{k}}\frac1{\sqrt{2\omega}}\left\{
a(\stackrel{\rightarrow}{k})e^{ik\cdot x}
+b^{\dag}(\stackrel{\rightarrow}{k})e^{-ik\cdot x}
\right\}
\eqno(4.1)  $$
has the invariance of ${\cal PT}={\cal C}$  evidently, i.e.,
when $x\rightarrow -x,\
e^{ik\cdot x}\rightarrow e^{-ik\cdot x},\ a(\stackrel{\rightarrow}{k})
\stackrel{\rightarrow}{\leftarrow}b^{\dag}
(\stackrel{\rightarrow}{k})$, one has
$$
\phi(x)\longrightarrow\phi_c(x)=\phi(x)\ .
\eqno(4.2)
$$
The situation for Dirac field is a little bit complicated:
$$
\psi(x)=
\frac1{\sqrt{V}}\sum_{\stackrel{\rightarrow}{p}}\sum_{r=1,2}\left\{
c_r(\stackrel{\rightarrow}{p})u^{(r)}(\stackrel{\rightarrow}{p})e^{ip\cdot x}
+d_r^{\dag}(\stackrel{\rightarrow}{p})v^{(r)}(\stackrel{\rightarrow}{p})e^{-ip\cdot x}
\right\}\ ,
\eqno(4.3)
$$
where $c_r (\stackrel{\rightarrow}{p})$ and $d^{\dag}
(\stackrel{\rightarrow}{p})$ are the annihilation operator of
electron and the
creation operator of positron respectively. For checking that (4.3) also
has the invariance of space-time inversion, we must also expand the
solution of Carid equation as a field operator:
$$
\psi_c(x)=
\frac1{\sqrt{V}}\sum_{\stackrel{\rightarrow}{p}}\sum_{r=1,2}\left\{
d_r^{\dag}(\stackrel{\rightarrow}{p})u^{(r)}(\stackrel{\rightarrow}{p})e^{-ip\cdot x}
+c_r(\stackrel{\rightarrow}{p})v^{(r)}(\stackrel{\rightarrow}{p})e^{ip\cdot x}
\right\}\ ,
\eqno(4.4)
$$
Then when $x\rightarrow -x,\ c_r(\stackrel{\rightarrow}{p})
\stackrel{\longrightarrow}{\leftarrow}d_r^{\dag}(\stackrel{\rightarrow}{p})$,
one has
$$
\psi(x)\longrightarrow\psi_c(x)=-\gamma_5\psi(x)\ .
\eqno(4.5)
$$

  So the ${\cal PT}  ={\cal C}$  invariance exhibit itself as the
transformation between
Dirac representation and Carid representation. From now on, we try to
propose an algorithm in quantum field theory: all field operators and the
Lagrangian density constructed from them, all the operator algebra,
should respect the invariance of ${\cal PT  =C}$ . In the following we
try to discuss
the relation between spin and statistics by means of this invariance.

  The spin-statistics connection was proved by Pauli [21] and discussed
by Schwinger and other authors [22, 23]. However, different kinds of proof
often carry some argument with negative character, among them the
violation of microcausality seems the strongest. For instance, the
complex K-G field is quantized via
$$
[\phi(x),\ \phi^{\dag}(y)]=i\hbar c\Delta(x-y)
\eqno(4.6)
$$
with $\Delta (-x)=-\Delta (x)$ and $\Delta (x)=0$ for $x^2>0$ (For
notation $\Delta (x)$, see Ref.[17]). The Dirac field is quantized via
 $$
\{\psi(x),\ \bar{\psi}(y)\}=-i\left(\gamma_{\mu}\frac{\partial}{\partial
x_{\mu}}
-\frac{mc}{\hbar}\right)\Delta(x-y)\ .
\eqno(4.7)
$$
The other function $\Delta_1(x)$ with property $\Delta_1  (-x)= \Delta_1
(x)$ is rejected because
$\Delta_1  (x)\neq 0$ for $x^2>  0$. Thus one obtains the correct statistics
and rejects the wrong ones.

  Note that the ${\cal PT}$   inversion does keep the Eq. (4.6)
invariant. It also
keep (4.7) invariant in the sense of transforming the latter into the
Carid representation:
 $$
\{\psi_c(x),\ \bar{\psi}_c(y)\}=-i\left(\gamma_{\mu}\frac{\partial}{\partial
x_{\mu}}
+\frac{mc}{\hbar}\right)\Delta(x-y)\ .
\eqno(4.8)
$$
On the very general ground we may replace the left hand side of (4.6) or
(4.7) by general bracket $[\ ,\ ]_{\omega}$   with $\omega =-1$
corresponding to commutation
while $\omega  =+1$ to anticommutation. Then a strong statement could be
that: `` Under
the condition of microcausality, ${\cal PT  = C}$  invariance determine the
correct statistics uniquely". At least we have a weak statement that: ``In
determining the spin statistics connection, the microcausality is in
conformity with the ${\cal PT  =C}$  invariance or vice versa". This
consistency seems
to us not quite a coincidence but has a deep implication as we will show
further in the sec. VI.

  We would like to point out that the antisymmetrical current which is
equivalent to the normal ordered current [17]:
$$
j_{\mu}=\frac12i[\bar{\psi}(x),\
\gamma_{\mu}\psi(x)]=i\stackrel{\cdot}{.}\bar{\psi}(x)
\gamma_{\mu}\psi(x)\stackrel{\cdot}{.}
\eqno(4.9)
$$
also has the transformation property
$$
{\cal C}j_{\mu}
{\cal C}^{-1}=-j_{\mu}^c=-j_{\mu}\ ,
\eqno(4.10)
$$
where the current operator at the right hand side is written in Carid
representation which in turn is equivalent to that in Dirac
representation. By this way, the special demand that all Lagrangians are
antisymmetrized in the fermion fields and symmetrized in the boson fields
can also be substituted by the unified requirement of ${\cal PT   =C}  $
invariance.

\vskip.2in
\leftline{{\Large \bf V. FEYNMAN PROPAGATOR}}
\vskip.2in
\leftline{{\Large \bf AND THE ARROW OF TIME}}
\vskip.2in
  Carrying on the concept in Newtonian mechanics, in quantum mechanics
there is a time reversal (T) transformation. However, because the
Schr\"odinger equation has the time derivative of first order, it changes
sign under $t\rightarrow -t$ transformation. So a further complex conjugate
transformation has to be made to cancel the sign change, which in turn
implies some equivalence between $\psi$ and $\psi^*$. The whole theory of time
reversal in quantum mechanics seems to us rather artificial and deserves
to be doubted.

  Actually, there is time asymmetry rather than symmetry in quantum
mechanics.
This can obviously be seen from the Feynman path integral formalism:
$$
\psi(\stackrel{\rightarrow}{x},\ t)=\int
 K(\stackrel{\rightarrow}{x},\ t|\stackrel{\rightarrow}{x'},\ t')
\psi(\stackrel{\rightarrow}{x'},\ t')d\stackrel{\rightarrow}{x'}\ ,\
\eqno(5.1)
$$
$$
 K(\stackrel{\rightarrow}{x},\ t|\stackrel{\rightarrow}{x'},\ t')
=\int_{\Gamma}{\cal D}\stackrel{\rightarrow}{x}e^{iS/\hbar}\ .
\eqno(5.2)
$$
In the expression of kernel $K$, the path $\Gamma$  takes every zigzag way
leading
from point $(\stackrel{\rightarrow}{x'},\ t')$ to
$(\stackrel{\rightarrow}{x},\
t)$ but without the reversal in time direction.
Alternatively, we may look at the corresponding Green function for
Schr\"odinger Equation:
$$
\left(i\hbar\frac{\partial}{\partial t}-H\right)
G(\stackrel{\rightarrow}{x},\ t|\stackrel{\rightarrow}{x'},\ t')
=
\delta (\stackrel{\rightarrow}{x}-\stackrel{\rightarrow}{x'})
\delta (t-t')\ .
\eqno(5.3)
$$
Then
$$
G(\stackrel{\rightarrow}{x},\ t|\stackrel{\rightarrow}{x'},\ t')
=
-\frac{i}{\hbar}
K(\stackrel{\rightarrow}{x},\ t|\stackrel{\rightarrow}{x'},\ t')
\theta(t-t')
$$
can be expanded by eigenfunctions of H as follows:
$$
G(\stackrel{\rightarrow}{x},\ t|\stackrel{\rightarrow}{x'},\ t')
=-\frac{i}{\hbar}
\sum_n
\phi_n(\stackrel{\rightarrow}{x})\phi_n^*(\stackrel{\rightarrow}{x'})\exp\{-\frac{i}{\hbar}E_n(t-t')\}
\theta(t-t')\ .
\eqno(5.4)
$$
The existence of $\theta (t-t')$ reflects the time asymmetry.

  In relativistic quantum mechanics, let us look at the Feynman
propagator $K_F$   for Dirac equation: [10]
$$
\left(\gamma_{\mu}\frac{\partial}{\partial x_{\mu}}+m\right)K_F(x,\
\stackrel{\rightarrow}{x})=-i\delta^{(4)}(x-x')\ ,\
\eqno(5.5)
$$
$$
K_F(x,\ x')=\sum_{\stackrel{\rightarrow}{p},\ s}\frac{m}{EV}\left\{
u^{(s)}(\stackrel{\rightarrow}{p})
\bar{u}^{(s)}(\stackrel{\rightarrow}{p})e^{ip\cdot(x-x')}
\theta(t-t')
-
v^{(s)}(\stackrel{\rightarrow}{p})
\bar{v}^{(s)}(\stackrel{\rightarrow}{p})e^{-ip\cdot(x-x')}
\theta(t'-t)\right\}\ .
\eqno(5.6)
$$

  It is still time asymmetric. But what new symmetry is there? Let us
perform a ${\cal PT}$   inversion on (5.5) and (5.6), we get
$$
\left(-\gamma_{\mu}\frac{\partial}{\partial x_{\mu}}+m\right)K_F^c(x,\
\stackrel{\rightarrow}{x})=-i\delta^{(4)}(x-x')\ ,\
\eqno(5.7)
$$
and
$$
K_F^c(x,\ x')=\sum_{\stackrel{\rightarrow}{p},\ s}\frac{m}{EV}\left\{
u^{(s)}(\stackrel{\rightarrow}{p})
\bar{u}^{(s)}(\stackrel{\rightarrow}{p})e^{-ip\cdot(x-x')}
\theta(t'-t)
-
v^{(s)}(\stackrel{\rightarrow}{p})
\bar{v}^{(s)}(\stackrel{\rightarrow}{p})e^{ip\cdot(x-x')}
\theta(t-t')\right\}\ .
\eqno(5.8)
$$
respectively. Notice that $K_F^c= \gamma_5K_F(x,\ x')\gamma_5$. So
except for the
representation transformation, (Dirac $\rightarrow$ Carid), $K_F^c(x,\
x')$         is essentially
the same as $K_F(x,\ x')$. In other words, it is the ${\cal PT=C} $
invariance that
forms the basis of selecting the Feynman propagator instead of simple
advanced or retarded one. This observation was put forward in Ref.[24],
where the relation between ${\cal PT}$  (i.e., CPT)invariance and the arrow of
time is discussed.

  At the microscopic level, the t asymmetry, i.e., the arrow of time is
dictated by a larger symmetry, the   ${\cal PT=C}$      invariance. While at
the level
of classical electrodynamics,        ${\cal PT=C}$ =CPT     invariance can
still play
the role of excluding the symmetric potential and selecting the correct
retarded potential in the sense of calculating the energy radiation
rate [24]. But the problem of how the macroscopic arrow of time, in
the sense of thermodynamics, will be related to the microscopic arrow of
time, is still controversial. It is well known that the H-theorem (i.e.,
the entropy increase principle) can only be proved in statistical
mechanics after a coarse grain density is defined in the phase space. Van
Hove generalized this theorem for quantum statistical case by defining a
coarse density matrix [25].
But what is the meaning of performing an averaging procedure for
defining the coarse grain density (matrix)? We conjecture that
the averaging procedure corresponds
to some operation which washes out the information of quantum phase and
then destroys the quantum coherence. Some preliminary discussion was
made via a simple system composed of two level atoms, radiations and
thermal reservoirs [26].
\vskip.2in
\leftline{{\Large \bf VI. THE SPECIAL RELATIVITY}}
\vskip.2in
\leftline{{\large \bf A. Where is the crucial point?}}
\vskip.2in

  We are now in a position to search for a new derivation of the theory of
special relativity. In his classical paper, Einstein introduced two
postulates: (A). the principle of the constancy of the speed of light;
(B). the principle of relativity. Since 1905 till recent years, some authors
felt that there is some logic cycle in these two relativistic
postulates, so they tried to derive the theory of special relativity
without the postulate (A). At final they all failed to do so. In
fact, Einstein had considered the problem from all aspects. While
talking about the principle of relativity, one needs the definition
of coordinates $\{x\}$ and $\{x'\}$ of two inertial systems, say $S$ and
$S'$, moving with relative velocity $\stackrel{\rightarrow}{v}$. The Lorentz
transformation is no
more than the definition of $\{x'\}$ with respect to $\{x\}$ or vice
versa. So
the constant $c$ in Lorentz transformation must be fixed in advance not
only in meaning but also in magnitude, otherwise one will have no
real physics in the principle of relativity. Einstein did the best
work in his time. He was the first physicist who emphasized the
necessity of distinguishing the observables from nonobservables.

  Now we are living after the quantum theory and particle physics all have
been well established. Can we derive the theory of special relativity by
only one "relativistic" postulate rather than two? Then it is clear that
we need a basic postulate stated only in one inertial system (S). In our
opinion, this postulate is nothing but the  ${\cal PT=C} $  symmetry
discussed in previous sections.
\vskip.2in
\leftline{{\large\bf B. The relativistic wave equation for spin zero
particle }}
\leftline{{\large\bf and the mass- energy relation}}
\vskip.2in
  We will approach this problem by considering the special cases
individually to see the role played by the   ${\cal PT=C}$  symmetry. In some
sense,
we will go along the opposite way in sec. III. Of course, in each case the
special postulate of nonrelativistic nature should be supplemented.

  Consider spin zero case first. A particle is resting in an inertial
system S. Assume its energy $E_0$ is proportional to its rest mass $m_0 ,\ E_0
= m_0 c_1^2 ,$ where $c_1$ being merely a constant with the dimension of
velocity.
Notice that, however, this is not an independent postulate (input),
because we will soon derive in general $E=mc^2_1 $, with mass $m$ defined as
$m=\frac12(\frac{dp^2}{dE}     )$.
Indeed, when particle is in slow motion, its
energy reads:
$$
E=m_0c_1^2+\frac{p^2}{2m_0},\ (\stackrel{\rightarrow}{p}\rightarrow0)\ .
\eqno(6.1)
$$
The rest mass does obey the relation:
$$
m_0=\frac12(\frac{dp^2}{dE}     )\ ,\
(\stackrel{\rightarrow}{p}\rightarrow0)\ .
 \eqno(6.2)
$$
The particle velocity $v$ equals to the group velocity of de Broglie wave
associated with it:
$$
v=v_g=\frac{d\omega}{dk}=\frac{dE}{dp}=\frac{p}{m_0},\ (v\rightarrow0)\ ,
\eqno(6.3)
$$
where the general quantum relations $E=\hbar\omega\ ,\   p=\hbar k$   and (6.1)
have been used.

  Consider that the wave is described by $
\theta (\stackrel{\rightarrow}{x},\ t)
$ and obeys the
following ``nonrelativistic quantum equation"
$$
i\hbar\frac{\partial}{\partial t}
\theta (\stackrel{\rightarrow}{x},\ t)
=m_0c_1^2\theta (\stackrel{\rightarrow}{x},\ t)
-\frac{\hbar^2}{2m_0}\nabla^2\theta (\stackrel{\rightarrow}{x},\ t)
\ ,\
\eqno(6.4)
$$
where the quantum rules
$$
E\longrightarrow i\hbar\frac{\partial}{\partial t},\
\stackrel{\rightarrow}{p}\longrightarrow-i\hbar\nabla \eqno(6.5)
$$
have been used. It is just at this moment we introduce the ``relativistic
principle" into the formula. That is   ${\cal PT=C}$  symmetry.
Corresponding to the
particle state $\theta(\stackrel{\rightarrow}{x},\ t)$ , there is an
antiparticle state $\chi (\stackrel{\rightarrow}{x}, t)$ hiding inside a
particle.   and   will couple to each other via the motion (kinetic
energy). So instead of (6.4), we should have a couple of equations:
$$
\left\{\begin{array}{l}
{\displaystyle
i\hbar\frac{\partial}{\partial t}
\theta
=m_0c_1^2\theta
-\frac{\hbar^2}{2m_0}\nabla^2\theta
-\frac{\hbar^2}{2m_0}\nabla^2\chi\ }.
\cr
{\displaystyle
i\hbar\frac{\partial}{\partial t}
\chi
=-m_0c_1^2\chi
+\frac{\hbar^2}{2m_0}\nabla^2\chi
+\frac{\hbar^2}{2m_0}\nabla^2\theta\ }.
\end{array} \right.
\eqno(6.6)
$$
Now Eqs. (6.6) respect the   ${\cal PT=C}$  symmetry because
$$
\chi(\stackrel{\rightarrow}{x},\ t)
=\theta(-\stackrel{\rightarrow}{x},\ -t)\ .
\eqno(6.7)
$$
Let $\theta =\frac12(\phi+i\xi)$       with $\phi$  and $\xi$  being real
functions of $(\stackrel{\rightarrow}{x},\ t)$
and notice from (6.6) that $\chi (\stackrel{\rightarrow}{x},\
t)=\theta^*(\stackrel{\rightarrow}{x},\ t)=\frac12(\phi-i\xi)$
we have
$$
\left\{\begin{array}{l}
\hbar\dot{\phi}=m_0c_1^2\xi\ ,
\cr
\hbar\dot{\xi}=-m_0c_1^2\phi+\frac{\hbar^2}{m_0}\nabla^2\phi\ .
\end{array}\right.
\eqno(6.8)
$$
Then the Klein-Gordon equation follows immediately:
$$
\left(\frac1{c^2}\frac{\partial^2}{\partial
t^2}-\nabla^2+\frac{m_0^2c_1^2}{\hbar^2}\right)
\phi(\stackrel{\rightarrow}{x},\ t)=0\ .
\eqno(6.9)
$$
Generalizing to the case of complex $\phi$  field, we set from the beginning
that
$$
\left\{\begin{array}{l}
\theta=\frac12(\phi+i\frac{\hbar}{m_0c_1^2}\dot{\phi})
\cr
\chi=\frac12(\phi-i\frac{\hbar}{m_0c_1^2}\dot{\phi})
\end{array}\right.
\eqno(6.10)
$$
and still get the K-G Eq. (6.9). Note also that $\theta\ ,\ \chi$  and
$\xi$  all satisfy K-G
equation. $|\chi|$    will increase from zero to $|\theta|$    when the
particle energy $E$ increases from $m_0c_1^2$   to infinity.

  Substituting the plane wave solution
$$
\phi(\stackrel{\rightarrow}{x},t)=\exp [i(\stackrel{\rightarrow}{p}\cdot
\stackrel{\rightarrow}{x}-Et)/\hbar]
\eqno(6.11)
$$
into (6.9), one obtains easily
$$
E^2=\stackrel{\rightarrow}{p}^2c_1^2+m_0^2c_1^4\ .
\eqno(6.12)
$$
Now the implication of constant $c_1$ has not been explored yet. The
velocity of
particle corresponds to the group velocity as in (6.3):
$$
v
=\frac{d\omega}{dk}
=\frac{dE}{dp}
=\frac{pc_1^2}{E}\ ,\ (|\stackrel{\rightarrow}{p}|=p)\ .
\eqno(6.13)
$$
Define the inertial mass as in (6.2):
$$
m=\frac{p}{v}
=p/\frac{dE}{dp}
=\frac12\frac{dp^2}{dE}\ .
\eqno(6.14)
$$
Combining (6.13), (6.14) and (6.12), one obtains
$$
E=mc_1^2    \eqno(6.15)
$$
and
$$
m=m_0\left(1-\frac{v^2}{c_1^2}\right)^{-\frac12}
\eqno(6.16)
$$
as expected. Evidently, $c_1$  is nothing but the limiting speed of
particle.
According to the experiments on $\pi$   meson beam, we know that $c_1$
equals to the speed of light, i.e.,
$$
c_1=c=3\times 10^{10} \ cm/sec\ .
\eqno(6.17)
$$
  In some sense, from $E|_{v\rightarrow0}=m_0c_1^2$           to $E=mc_1^2$, we
are
using a trick similar
to the inductive method in mathematics. But the ``relativistic hormone" is
just the symmetry    ${\cal PT=C}$.
\vskip.2in
\leftline{ {\large \bf C. The relativistic wave equations for spin
1/2}}
\vskip.2in
  The present experiment and theory reveal that the neutrino is likely
a particle with spin 1/2 and zero rest mass. So its wave equation may
be
$$
E\phi_R=i\hbar\frac{\partial}{\partial
t}\phi_R=c_2\stackrel{\rightarrow}{p}\cdot\stackrel{\rightarrow}{\sigma}\phi_R=-ic_2\stackrel{\rightarrow}{\sigma}\cdot\nabla\phi_R
\eqno(6.18)
$$
or
$$
E\phi_L=i\hbar\frac{\partial}{\partial
t}\phi_L=-c_2\stackrel{\rightarrow}{p}
\cdot\stackrel{\rightarrow}{\sigma}\phi_L
=ic_2\stackrel{\rightarrow}{\sigma}\cdot\nabla\phi_L
\eqno(6.19)
$$
As before, here $c_2$  is still an unfixed constant with dimension of
velocity. Both these two Weyl equations respect the    ${\cal PT=C}$
symmetry,
but the nature seems to favor (6.19) and discard (6.18), as we learn
from the two-component neutrino theory.

  Now consider a particle with spin 1/2 and rest mass $m_0$. Then
two-component spinors $\phi_L$  and $\phi_R$  will couple each other via
$m_0$  (rather than via the ``kinetic energy
$c_2\stackrel{\rightarrow}{p}\cdot \stackrel{\rightarrow}{\sigma}$ "):
 $$
\left\{\begin{array}{l}
i\hbar\frac{\partial}{\partial t}\phi_R
=-ic_2\stackrel{\rightarrow}{\sigma}\cdot\nabla\phi_R+m_0c_2^2\phi_L \ ,
\cr
i\hbar\frac{\partial}{\partial t}\phi_L
=ic_2\stackrel{\rightarrow}{\sigma}\cdot\nabla\phi_L+m_0c_2^2\phi_R \ .
\end{array}\right.
\eqno(6.20)
$$
Defining four-component spinor
$$
\psi=
\left(\begin{array}{l}
\theta
\cr
\chi
\end{array}\right)
=
\left(\begin{array}{l}
\phi_R+\phi_L
\cr
\phi_R-\phi_L
\end{array}\right) \ .
\eqno(6.21)
$$
one recovers the Dirac equation again[10]:
$$
\left(\gamma_{\mu}\frac{\partial}{\partial
x_{\mu}}+\frac{m_0c_2}{\hbar}\right)\psi=0\ .
\eqno(6.22) $$
The equation obeyed by $\theta$   and $\chi$   in external fields had been
written in
(3.8). The experiments on electron beam verify that the limiting speed
$c_2 =c=3\times10^{10}\ cm/sec.$

  The important thing here is never looking at a wave function (like
$\psi$ )
as a whole entity but an object composed of two parts in contradiction (
$\theta$ and $\chi$, or $\phi_R$  and $\phi_L$), while respecting
the   ${\cal PT=C}$     symmetry at the same time.

  The similar experience used to spin 1 particle, photon, had been
written in Eqs.(3.15)-(3.21). This time the wave function
$\Phi=(\stackrel{\rightarrow}{E}+i\stackrel{\rightarrow}{B})$         is
composed of the real and imaginary parts with $\stackrel{\rightarrow}{E}$   and
$\stackrel{\rightarrow}{B}$   being
observables.
\vskip.2in
\leftline{{\large \bf D.The Lorentz transformation.}}
\vskip.2in
  As explained at the beginning of this section, in Einstein's theory,
the principle of the constancy of light speed must be put ahead of the
principle of relativity. Now we are going to derive the Lorentz
transformation after some knowledge about the dynamics of particles is
known. Nevertheless, we still need an invariance to define the
coordinates $\{x\}$ and $\{x'\}$ in $S$ and $S'$ systems. Once again, we
resort to the
brilliant idea of de Broglie, who named it as ``the law of phase harmony" and
was explained by Lochak as follows [27]. For any Galilean observer, the
phase of the ``internal clock" of the particle is, at each instant, equal
to the value of the phase of the wave calculated at the same point at
which the particle lies.
(de Broglie considered this law to be the fundamental achievement in
his life and it was appreciated very much by Einstein. However, it was
nearly forgotten in most text books.).

  The phase of wave reads
$(\stackrel{\rightarrow}{p}\cdot \stackrel{\rightarrow}{x}-Et)/\hbar  =
(\stackrel{\rightarrow}{k}\cdot \stackrel{\rightarrow}{x}  -\omega t  )$,
while the phase of
``internal clock" (a stationary wave associated with the particle) will
be $-E t'/\hbar  = -\omega_0 t'$. The equality means:
$$
(\stackrel{\rightarrow}{p}\cdot \stackrel{\rightarrow}{x}-Et)  =
-E_0t'\ .
\eqno(6.23)
$$
Consider that the momentum $p=mv$ is along the $x$ axis. Substituting the
expressions for $\stackrel{\rightarrow}{p},\ E$ and  $E_0$  into (6.23),
one finds
$$
t'=\left(t-\frac{v}{c^2}x\right)
\left(1-\frac{v^2}{c^2}\right)^{-\frac12}\ .
\eqno(6.24)
$$
Now our laboratory is $S$ system, while $S'$ system resting on particle is
moving with velocity $v$. The clock is located at the origin of $S'$ system,
so $x'=0$ and a generic point at $S'$ system will have the coordinate
$$
x'=a(x-vt)\ ,           \eqno(6.25)
$$
where $a$ is a constant. Because the difference between $S$ and $S'$ is
relative in direction, we anticipate
$$
t=\left(t'+\frac{v}{c^2}x\right)
\left(1-\frac{v^2}{c^2}\right)^{-\frac12}\ .
\eqno(6.26)
$$
Substituting (6.24) and (6.25) into (6.26), one finds
$$
a=\left(1-\frac{v^2}{c^2}\right)^{-\frac12}\ .
\eqno(6.27)
$$
Alternatively, one may use more abstract language. The phase, i.e. ,the
right hand side of (6.23) is an invariant in the sense of being a
constant with respect to the change of $\stackrel{\rightarrow}{v}$. On the
other hand, we have
already
$$
\stackrel{\rightarrow}{p}^2+\left(\frac{iE}{c}\right)^2=-m_0^2c^2=const.\ .
\eqno(6.28)
$$
Rewriting (6.23)in the form as
$$
 \stackrel{\rightarrow}{p}\cdot
\stackrel{\rightarrow}{x}+\left(\frac{iE}{c}\right)(ict)=const. \ ,
\eqno(6.29) $$
we see that there is a  four dimensional vector
$(\stackrel{\rightarrow}{p},\ iE/c)$ in Minkowski
space with its length  fixed. This vector and
$(\stackrel{\rightarrow}{x},\ ict)$ construct a scalar
product which is also
invariant with respect to the transformation of coordinates between $S$ and
$S'$. This means that $(\stackrel{\rightarrow}{x},\ ict)$ is also a four
vector in Minkowski space with
its length fixed. To be precise, the four dimensional interval between
two space-time points is an invariant:
$$
(\Delta \stackrel{\rightarrow}{x})^2+(ic\Delta t)^2
=(\stackrel{\rightarrow}{x}_1-\stackrel{\rightarrow}{x}_2)^2+[ic(t_1-t_2)]^2
=(\stackrel{\rightarrow}{x'}_1
-\stackrel{\rightarrow}{x'}_2)^2+[ic(t'_1-t'_2)]^2 =const. \ . \eqno(6.30)
$$
This equation together with (6.24) uniquely determines the Lorentz
transformation:
$$
\begin{array}{ll}
\left\{
\begin{array}{l}
\displaystyle{
x'=\frac{x-vt}{\sqrt{1-\frac{v^2}{c^2}}} }
\cr
\displaystyle{
t'=\frac{t-\frac{v}{c^2}x}{\sqrt{1-\frac{v^2}{c^2}}}}
\cr
y'=y,\ z'=z
\end{array}\right.
&
\left\{
\begin{array}{l}
\displaystyle{
x=\frac{x'+vt'}{\sqrt{1-\frac{v^2}{c^2}}}}
\cr
\displaystyle{
t=\frac{t'+\frac{v}{c^2}x'}{\sqrt{1-\frac{v^2}{c^2}}}}
\cr
y=y',\ z=z'
\end{array}\right.
\end{array}
\eqno(6.31)
$$
  Note that the constant c has the meaning of limiting speed of the
particle, it equals to the speed of light experimentally. Moreover, the
time dilatation effect can now be understood
as follows. A particle state characterizing by $\theta
(\stackrel{\rightarrow}{x},\ t)$ is always
accompanying with some antiparticle ingredient characterizing by
$\chi(\stackrel{\rightarrow}{x},\ t)$ which has the opposite space-time phase
dependence
essentially(implicitly). The observer in $S$ system looks at the moving
"internal clock" with its phase change as the time record. It seems
slower and slower as the magnitude of $\chi$   increases larger and larger
together with the increase of velocity $v$. In the limiting case, $v\rightarrow
c,\ |\chi|\rightarrow|\theta|      ,$ the moving "clock" tends to stop and
its mass approaches to
infinity. So we find that the kinetic effect in special relativity does
have its universal dynamical origin. In some sense, the special
relativity is more quantum than quantum mechanics.
\vskip.2in
\leftline{{\Large\bf  VII. SUMMARY AND DISCUSSION.}}
\vskip.2in
(1) It is shown from sec. II  to sec. V  that the relativistic quantum
mechanics and quantum field theory all contain a basic
invariance$-$  ${\cal PT=C}$        symmetry $-$ explicitly or implicitly. In
particular,
the following four things are linked together via this postulate quite
naturally:

(a) The appearance of negative energy or negative probability density,
which corresponds to the fact of existence of antiparticle in single
particle equations;

(b) Spin-statistics theorem in many particle theory;

(c) CPT theorem;

(d) The Feynman propagator.

(2) Now a question arises. If the new postulate is used to replace the CPT
theorem, the number of input to our foundation of physics would be
increased. This could make the situation even worse in view of the
criterion$--$less input, more output$--$long established in theoretical
physics. So in sec. VI we devote to deriving the special relativity via
the only ``relativistic principle" in microscopic sense, i.e. , the  ${\cal
PT=C}$
postulate. Especially, we obtain:

(e) The mass-energy relation;

(f) The Lorentz transformation.

  Of course, we stress again, some well established "nonrelativistic"
knowledge (postulate) and/or special prescription for individual case are
needed.

  Therefore, as a whole, the number of input to our foundation of physics
is less ( at least no more) than before.

(3) Evidently, we are treating the wave function, especially its phase,
much more seriously than before. This is relevant to the basic
explanation of quantum mechanics and deserves further investigation.

(4) What we have done could be depicted by a diagram, see FIG.1 [28]. The
lower part of this diagram is unobservable. Once when a particle is
excited from the vacuum, it becomes observable. How can one detects a
particle in experiments? One is relying on its apparent
momentum or energy. So the momentum or energy
is the existence form of a
particle. In some sense we would also say that the space- time is the
existence form of vacuum.  There are two dotted lines connecting the
lower part to upper part. The left one implies the quantum operator rule:
$$
\stackrel{\rightarrow}{p}\longrightarrow-i\hbar\nabla\ ,\
\stackrel{\rightarrow}{p}_c\longrightarrow i\hbar\nabla\ ,\
\eqno(7.1)
$$
while the right one implies:
$$
E\longrightarrow i\hbar\frac{\partial}{\partial t}\ ,\
E_c\longrightarrow-i\hbar\frac{\partial}{\partial t}\ ,
\eqno(7.2)
$$
Now a universal constant, the Planck constant ($\hbar = h/2\pi$), emerges
as a
vertical link. On the other hand, the horizontal link on this diagram is
provided by an another universal constant, the light speed $c$.
Historically, Einstein discovered the horizontal link, first in the lower
part, then in the upper part. Only after the knowledge about the
content-- how the particles exhibit themselves as the excitation states
of vacuum-- together with the quantum theory (the vertical link) have
been accumulating in the past ninety years, can we try to examine this
diagram via some what different way from that of Einstein.
\vskip.2in
\leftline{{\large \bf ACKNOWLEDGEMENTS} }
\vskip.2in
This work was supported in part by the NSF in China. We also thank the
kind hospitality of ICTP, Trieste, Italy, where the manuscript was completed.

\break
\centerline{{\large \bf APPENDIX A. THE ORIENTATION OF SPIN}}

\vskip.3in
\begin{tabular}{l|l}

Dirac Equation
   &                  Carid Equation\\
$H\psi=i\hbar\frac{\partial\psi}{\partial t}$         &
$H_c\psi_c=-i\hbar\frac{\partial\psi_c}{\partial t}$  \\
$H=c\stackrel{\rightarrow}{\alpha}\cdot \stackrel{\rightarrow}{p} +\beta
mc^2,\ \stackrel{\rightarrow}{p}=-i\hbar\nabla $  &
$H_c=c\stackrel{\rightarrow}{\alpha}\cdot \stackrel{\rightarrow}{p}_c
+\beta mc^2,\ \stackrel{\rightarrow}{p}_c=i\hbar\nabla $ \\
A physical observable $F$ will  &
  A physical observable F will\\
change with time as             &
  change  with time as  \\
$\frac{dF}{dt}=
\frac{\partial F}{\partial t}+\frac{i}{\hbar}[H,\ F] $
 &
$\frac{dF}{dt}=
\frac{\partial F}{\partial t}-\frac{i}{\hbar}[H_c,\ F] $
\\
{}From the orbital angular  &
   From the orbital angular  \\
momentum of an electron                   &
    momentum of a positron        \\
$\stackrel{\rightarrow}{L}=\stackrel{\rightarrow}{x}\times
\stackrel{\rightarrow}{p} $
   &
$\stackrel{\rightarrow}{L}_c=\stackrel{\rightarrow}{x}\times
\stackrel{\rightarrow}{p}_c $
\\
and spin
 &                          and spin   \\
$ \stackrel{\rightarrow}{\Sigma}=\left(\begin{array}{cc}
\stackrel{\rightarrow}{\sigma} & 0 \\
0  &  \stackrel{\rightarrow}{\sigma} \end{array}\right) $            &
$ \stackrel{\rightarrow}{\Sigma}=\left(\begin{array}{cc}
\stackrel{\rightarrow}{\sigma} & 0 \\
0  &  \stackrel{\rightarrow}{\sigma} \end{array}\right)  $
\\
one has
&                             one has   \\
$\frac{d}{dt}(\stackrel{\rightarrow}{L}+\frac{\hbar}{2}\stackrel{\rightarrow}{\Sigma})=0 $      &
$\frac{d}{dt}(\stackrel{\rightarrow}{L}_c-\frac{\hbar}{2}\stackrel{\rightarrow}{\Sigma})=0 $
\\
This means the conservation of total  &
 This means the conservation of total     \\
angular momentum. The first and        &
 angular momentum. The first and            \\
third components of spinor              &
 third components of spinor                   \\
correspond to spin along the $z$ axis    &
 correspond to spin along ($-z$) axis           \\
(up, $\uparrow$) while the second and fourth &
 (down, $\downarrow$) while the second and fourth \\
ones to spin down ($\downarrow$) states of    &
 ones to spin up ($\uparrow$) states of  \\
electron.                            &
 positron. \\
For positron, the first and third  &
 For electron, the first and third \\
components imply spin down  ($\downarrow$) states &
state  components imply spin up ($\uparrow$) \\
while the second and fourth ones &
while the second and fourth ones               \\
being up ($\uparrow$) states.     &
being down ($\downarrow$) states. \\
\end{tabular}

 \break
\centerline{{\large\bf APPENDIX B.THE SOLUTIONS OF DIRAC EQUATION}}
\centerline{{\large\bf  AND
CARID EQUATION}}
\vskip.2in
\begin{tabular}{l|l}
     Dirac equation
    &                    Carid equation
\\
$\gamma_{\mu}\frac{\partial}{\partial
x_{\mu}}\psi(x)+\frac{mc}{\hbar}\psi(x)=0$  &
$\gamma_{\mu}\frac{\partial}{\partial
x_{\mu}}\psi_c(x)-\frac{mc}{\hbar}\psi_c(x)=0$  \\

There are four independent solutions & There are four independent solutions
\\

$\psi_I(x)=u^{(1)}(\stackrel{\rightarrow}{p})\exp\{i(\stackrel{\rightarrow}{p}\cdot \stackrel{\rightarrow}{x}-Et)/\hbar\}$ &

$\psi_{cI}(x)=u^{(1)}(\stackrel{\rightarrow}{p})\exp\{-i(\stackrel{\rightarrow}{p}\cdot \stackrel{\rightarrow}{x}-Et)/\hbar\}$ \\

$\psi_{II}(x)=u^{(2)}(\stackrel{\rightarrow}{p})\exp\{i(\stackrel{\rightarrow}{p}\cdot \stackrel{\rightarrow}{x}-Et)/\hbar\}$   &

$\psi_{cII}(x)=u^{(2)}(\stackrel{\rightarrow}{p})\exp\{-i(\stackrel{\rightarrow}{p}\cdot \stackrel{\rightarrow}{x}-Et)/\hbar\}$   \\

$\psi_{III}(x)=v^{(1)}(\stackrel{\rightarrow}{p})\exp\{-i(\stackrel{\rightarrow}{p}\cdot \stackrel{\rightarrow}{x}-Et)/\hbar\}$      &

$\psi_{cIII}(x)=v^{(1)}(\stackrel{\rightarrow}{p})\exp\{i(\stackrel{\rightarrow}{p}\cdot \stackrel{\rightarrow}{x}-Et)/\hbar\}$       \\

$\psi_{IV}(x)=v^{(2)}(\stackrel{\rightarrow}{p})\exp\{-i(\stackrel{\rightarrow}{p}\cdot \stackrel{\rightarrow}{x}-Et)/\hbar\}$   &

$\psi_{cIV}(x)=v^{(2)}(\stackrel{\rightarrow}{p})\exp\{i(\stackrel{\rightarrow}{p}\cdot \stackrel{\rightarrow}{x}-Et)/\hbar\}$     \\
The former two solutions describe  &2
 The former two solutions describe    \\
the electron, while latter two      &
 the positron, while latter two    \\
describe positron.                   &
 describe electron.
\end{tabular}
$$\begin{array}{cc}
u^{(1)}(\stackrel{\rightarrow}{p})=N\left(\begin{array}{c}
1\\
0\\
\displaystyle{\frac{p_3c}{E+mc^2}}\\
\displaystyle{\frac{(p_1+ip_2)c}{E+mc^2}}\end{array}
\right)\ ,\     &
u^{(2)}(\stackrel{\rightarrow}{p})=N\left(\begin{array}{c}
0\\
1\\
\displaystyle{\frac{(p_1-ip_2)c}{E+mc^2}}\\
\displaystyle{\frac{-p_3c}{E+mc^2}}
\end{array}
\right)\ ,  \\
v^{(1)}(\stackrel{\rightarrow}{p})=N\left(\begin{array}{c}
\displaystyle{\frac{p_3c}{E+mc^2}}\\
\displaystyle{\frac{(p_1+ip_2)c}{E+mc^2}}  \\
1\\
0\\
\end{array}
\right)\ ,\     &
v^{(2)}(\stackrel{\rightarrow}{p})=N\left(\begin{array}{c}
\displaystyle{\frac{(p_1-ip_2)c}{E+mc^2}}\\
\displaystyle{\frac{-p_3c}{E+mc^2}}\\
0\\
1
\end{array}
\right) \ ,
\end{array}
$$

$$E=\sqrt{p^2c^2+m^2c^4}>0\ ,\ N=\sqrt{\frac{E+mc^2}{2mc^2}}\ , $$

$$u^{(r)^{\dag}}(\stackrel{\rightarrow}{p})u^{(r)}(\stackrel{\rightarrow}{p})=\frac{E}{mc^2},\ v^{(r)}(\stackrel{\rightarrow}{p})
=-\gamma_5u^{(r)}(\stackrel{\rightarrow}{p}),\ (r=1,\ 2).$$

 \break

\centerline{{\large\bf APPENDIX C. FURTHER DISCUSSION ON DIRAC
PARTICLES}}
\vskip.2in
  It is well known that the negative-energy solution become appreciable
during the compression of a electron packet and is responsible for the
phenomena of zitterbewegung [10, 13, 29]. The interpretation will be
inconsistent if the negative states are all filled as in the hole theory.

  Either in free case or in an external field, the positive and negative
solutions together constitute a complete set. Starting from this point, Ma
and Ni derived the Levinson theorem for Dirac particles [30]. In Eq. (20)
of Ref. [30], it was shown that when the particle is moving in a short
range attractive central potential $V(r)$ the phase shift is positive for
the positive-energy states and negative for the negative-energy states.
This implies that for the latter case the particle is repulsed by the
potential and so behaves as an antiparticle.

  In this paper, we show that it is the coherent excitation of
antiparticle ingredient inside a particle state which is responsible for
the generation of "kinetic mass" of the particle. Where the rest mass
comes from?

  Let us look at the mass-energy relation again:
$$
E^2=m^2c^4=m_0^2c^4+p^2c^2\ .               \eqno(C.1)
$$
Notice that in a composite particle , the "total mass" $m$   of a
constituent particle becomes part of the "rest mass" $m_0$     of composite
particle. This fact implies that the rest mass $m_0$     and the "kinetic
mass" ($p/c$) must be stemming from the same origin. However, the right
triangle relation between $m_0$    and  ($p/c$), Eq. (C.1), strongly hints
that they must be generated from different (orthogonal) mechanism.

  In Ref. [31], basing on NJL model [14c], we redrive the formula (C.1)
with $m_0$ being an energy gap in the new vacuum, which is formed after
the
condensation of massless particle-antiparticle pairs in original (naive)
vacuum. So we see that the rest mass and kinetic mass are generated via
different  orthogonal mechanism (many body effect versus single particle
effect), while both of them are stemming from the common origin$-$the
coexistence of particle-antiparticle and the  ${\cal PC=T}$
symmetry.

\break

\break

\centerline{{ \LARGE \bf Figure Caption}}
\vskip.3in
FIG.1. The special relativity and quantum mechanics are two pillars of
modern physics as viewed by the achievement of particle physics. However,
it is just the development of particle physics which reveals the
microscopic essence of special relativity. For more detail, see text.

\break

\begin{figure}

\begin{picture}(300,300)(-40,-90)
\put(0,0){\framebox(260,240)}
\put(50,130){\makebox(0,0){\footnotesize $\hbar$}}
\put(210,130){\makebox(0,0){\footnotesize $\hbar$}}
\put(130,60){\circle{80}}
\put(130,180){\circle{80}}
\put(140,145){\line(-1,1){10}}
\put(120,145){\line(1,1){10}}
\put(135,85){\line(0,1){65}}
\put(125,85){\line(0,1){65}}
\put(130,60){\makebox(0,0){\footnotesize Vacuum}}
\put(130,180){\makebox(0,0){\footnotesize Particle}}
\put(130,20){\makebox(0,0)  c}
\put(135,20){\vector(1,0){65}}
\put(125,20){\vector(-1,0){65}}
\put(130,220){\makebox(0,0) c}
\put(135,220){\vector(1,0){65}}
\put(125,220){\vector(-1,0){65}}
\put(60,60){\makebox(0,0){\footnotesize Space ($\stackrel{\rightarrow}{x}$)}}
\put(57,160){$\uparrow$}
\put(200,60){\makebox(0,0){\footnotesize Time (t)}}
\put(197,160){$\uparrow$}
\put(60,70){\line(0,1){2}}
\put(60,73){\line(0,1){2}}
\put(60,76){\line(0,1){2}}
\put(60,79){\line(0,1){2}}
\put(60,82){\line(0,1){2}}
\put(60,85){\line(0,1){2}}
\put(60,88){\line(0,1){2}}
\put(60,91){\line(0,1){2}}
\put(60,94){\line(0,1){2}}
\put(60,97){\line(0,1){2}}
\put(60,100){\line(0,1){2}}
\put(60,103){\line(0,1){2}}
\put(60,106){\line(0,1){2}}
\put(60,109){\line(0,1){2}}
\put(60,112){\line(0,1){2}}
\put(60,115){\line(0,1){2}}
\put(60,118){\line(0,1){2}}
\put(60,121){\line(0,1){2}}
\put(60,124){\line(0,1){2}}
\put(60,127){\line(0,1){2}}
\put(60,130){\line(0,1){2}}
\put(60,133){\line(0,1){2}}
\put(60,136){\line(0,1){2}}
\put(60,139){\line(0,1){2}}
\put(60,142){\line(0,1){2}}
\put(60,145){\line(0,1){2}}
\put(60,148){\line(0,1){2}}
\put(60,151){\line(0,1){2}}
\put(60,154){\line(0,1){2}}
\put(200,70){\line(0,1){2}}
\put(200,73){\line(0,1){2}}
\put(200,76){\line(0,1){2}}
\put(200,79){\line(0,1){2}}
\put(200,82){\line(0,1){2}}
\put(200,85){\line(0,1){2}}
\put(200,88){\line(0,1){2}}
\put(200,91){\line(0,1){2}}
\put(200,94){\line(0,1){2}}
\put(200,97){\line(0,1){2}}
\put(200,100){\line(0,1){2}}
\put(200,103){\line(0,1){2}}
\put(200,106){\line(0,1){2}}
\put(200,109){\line(0,1){2}}
\put(200,112){\line(0,1){2}}
\put(200,115){\line(0,1){2}}
\put(200,118){\line(0,1){2}}
\put(200,121){\line(0,1){2}}
\put(200,124){\line(0,1){2}}
\put(200,127){\line(0,1){2}}
\put(200,130){\line(0,1){2}}
\put(200,133){\line(0,1){2}}
\put(200,136){\line(0,1){2}}
\put(200,139){\line(0,1){2}}
\put(200,142){\line(0,1){2}}
\put(200,145){\line(0,1){2}}
\put(200,148){\line(0,1){2}}
\put(200,151){\line(0,1){2}}
\put(200,154){\line(0,1){2}}
\put(60,180){\makebox(0,0){\footnotesize Momentum
($\stackrel{\rightarrow}{p}$)}}
\put(200,180){\makebox(0,0){\footnotesize Energy (E)}}
\put(0,120){\line(1,0){260}}
\put(0,-55){\makebox(0,0){\footnotesize FIG. 1. }}
\end{picture}
\end{figure}

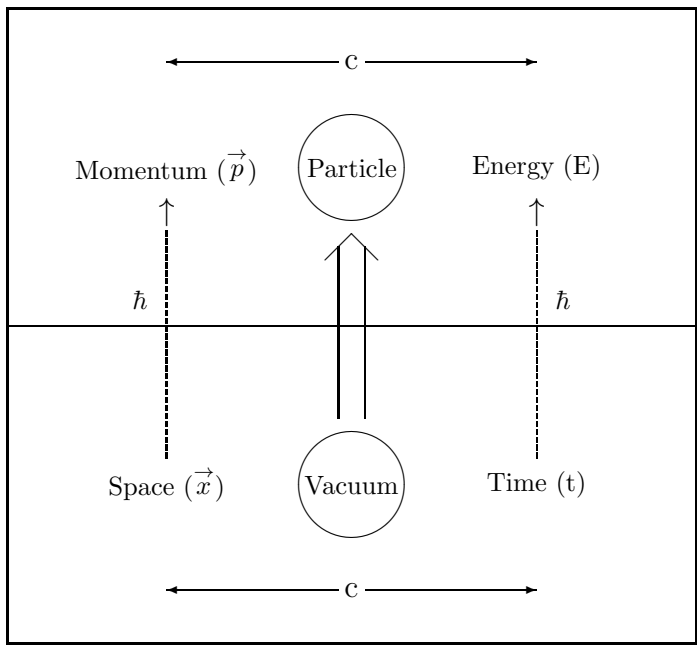
\end{document}